# Using the Conceptual Survey of Electricity and Magnetism to investigate progression in student understanding from introductory to advanced levels

Alexandru Maries[1], Mary Jane Brundage[2] and Chandralekha Singh[2]

[1] Department of Physics, University of Cincinnati, Cincinnati, OH, 45221

[2] Department of Physics and Astronomy, University of Pittsburgh, Pittsburgh, PA, 15260

**Abstract.** The Conceptual Survey of Electricity and Magnetism (CSEM) is a multiple-choice survey that contains a variety of electricity and magnetism concepts from Coulomb's law to Faraday's law at the level of introductory physics used to help inform instructors of student mastery of those concepts. Prior studies suggest that many concepts on the survey are challenging for introductory physics students and the average student scores after traditional instruction are low. The research presented here investigates the progression in student understanding on the CSEM. We compare the performance of students in introductory and advanced level physics courses to understand the evolution of student understanding of concepts covered in the CSEM after traditional lecture-based instruction. We find that on all CSEM questions on which less than 50% of the introductory physics students answered a question correctly after instruction, less than two thirds of the upper-level undergraduate students provided the correct response after traditional instruction. We also analyzed the CSEM data from graduate students for benchmarking purposes. We discuss the CSEM questions that remain challenging and the common alternative conceptions among upper-level students. The findings presented here at least partly point to the fact that traditional instruction in upper-level courses which typically focuses primarily on quantitative problem solving and incentivizes use of algorithmic approaches is not effective for helping students develop a solid understanding of these concepts. However, it is important for helping students integrate conceptual and quantitative aspects of learning in order to build a robust knowledge structure of basic concepts in electricity and magnetism.

## INTRODUCTION

One goal of an introductory electricity and magnetism (EM) course for physical science and engineering majors is for students to develop a robust conceptual understanding of the underlying physics. For example, in the context of circuits, University of Washington physics education group has developed several tutorials [1, 2], Eylon and Ganiel have investigated macro-micro relationships to analyze the link between electrostatics and electrodynamics [3], Thacker et al. have investigated student understanding of transients in direct current circuits [4], and Li and Singh have focused on student common difficulties and approaches while solving conceptual problems with non-identical light bulbs in series and parallel [5]. Other prior investigations in EM have focused on investigating student conceptual understanding of Coulomb's law, superposition principle, electric flux and Gauss' law [6-11], capacitance [12, 13], and Faraday's law and electromagnetic induction [14, 15]. Furthermore, prior research has also focused on assessing student problem solving skills in the context of introductory electricity and magnetism courses [16-21] and improving understanding of these challenging concepts using evidence-based active-learning approaches [22-24].

Improving student conceptual understanding and problem solving remain major goals in courses for physics majors as well as for physics graduate students as they take advanced physics courses even though these courses use complex mathematics to solve problems [25-28]. In these advanced courses, students are often expected to make the math-physics connection themselves to develop a good knowledge structure without much guidance and support or incentive from instructors to integrate conceptual and quantitative aspects of the material [29, 30]. In particular, the focus of traditionally taught upper-level undergraduate and graduate courses is often on being able to solve complex quantitative problems that can be solved using algorithmic approaches with course assessment focusing exclusively on such problems, which students can solve without having developed a robust understanding of the underlying concepts.

In the context of EM, students often learn about the basic concepts including Maxwell's equations in integral form in the calculus-based introductory courses. However, in the typical traditionally taught upper-level undergraduate and graduate EM courses, students are taught to use differential forms of the Maxwell's equations to solve complex quantitative problems that can be solved algorithmically often without any course assessment focusing specifically on conceptual understanding. If advanced students in these traditionally taught courses are not given grade incentives and provided scaffolding support to make the appropriate math-physics connections, they may solve problems by pattern matching without necessarily unpacking the underlying concepts for developing a deep conceptual understanding of the foundational concepts, which is important for cultivating expertise in physics [29,

30]. For example, the "cognitive apprenticeship" model [31] is a field-tested model for helping students develop expertise and emphasizes that not only should instructors model effective approaches to learning and problem solving but provide students with coaching and scaffolding support with a gradual weaning of the support as students develop self-reliance. Providing "scaffolding support" is one of the important aspects of cognitive apprenticeship model [31] defined by the authors as, "*Scaffolding* refers to the supports the teacher provides to help the student carry out a task. [Scaffolding] involves a kind of cooperative problem-solving effort by teacher and student in which the express intention is for the student to assume as much of the task on his own as possible, as soon as possible." Lack of scaffolding support can result in even advanced physics students not being able to integrate the conceptual and quantitative aspects of physics which is essential for developing expertise.

Many conceptual multiple-choice assessments have been developed and validated to evaluate student understanding of basic concepts covered in introductory physics [32-42]. One such survey is The Conceptual Survey of Electricity and Magnetism or CSEM [33], which has been widely used to investigate the conceptual understanding of students in introductory EM courses before and after instruction in relevant concepts. It is a 32-question multiple-choice survey in which EM topics range from Coulomb's law to Faraday's law. The CSEM can be used to measure the effectiveness of evidence-based curricula and pedagogies. Past research using other surveys has shown that even in upper-level undergraduate classes, students have difficulty with many introductory and advanced level concepts [43, 44].

## THEORETICAL FRAMEWORK, MOTIVATION AND GOAL FOR THIS STUDY

This study is motivated by the theoretical framework that as students progress from introductory to advanced courses in a particular area of physics, e.g., EM, each advanced course should help them develop a deeper understanding of the underlying concepts even though the mathematical sophistication of the course material increases. The study reported here uses the CSEM to investigate the progression in student conceptual understanding after traditional lecture-based instruction from introductory to advanced levels. The use of conceptual validated surveys such as CSEM can be useful for investigating the extent to which traditionally taught courses are succeeding in improving student conceptual understanding as students progress from introductory to advanced level courses. Using conceptual surveys can help benchmark this evolution and compare it with future studies involving evidence-based physics courses in which there is an explicit focus on integrating conceptual and quantitative aspects of learning.

This study is valuable for starting a discourse about the extent to which upper-level courses taught traditionally, e.g., the EM course discussed here, are helping students develop a better understanding of foundational EM concepts and the need for upper-level instructional design to intentionally integrate conceptual and quantitative aspects of learning so that upper-level students develop a robust knowledge structure of these basic concepts. Furthermore, this study is useful not only for instruction in upper-level EM courses but also in introductory courses because knowledge about the concepts that are challenging for both populations can help instructors contemplate why those concepts continue to be challenging and develop effective evidence-based approaches for improving student conceptual understanding of EM throughout.

We compared introductory students' performance at the end of a calculus-based introductory physics course with that of upper-level undergraduates at the end of an upper-level EM course. We also administered the CSEM to first-year physics graduate students before their graduate core electricity and magnetism course to benchmark the performance of undergraduates. A central goal of the research was to investigate the extent to which a traditionally taught upper-level undergraduate EM course, which primarily focuses on quantitative problem solving with regard to teaching and assessment, helped students develop conceptual understanding of the introductory concepts covered in the CSEM. We focused on identifying concepts which upper-level students still had difficulties with at the end of a traditionally taught upper-level EM course and analyzed different patterns of evolution across the CSEM questions from the introductory to the upper-level, e.g., the types of questions on which advanced students improved significantly or did not do so. In particular, we investigated the CSEM concepts that were challenging for both introductory and upper-level students and the CSEM concepts that were not mastered at the introductory level but were mastered at the upper-level. Additionally, we analyzed data to investigate if there are CSEM questions on which the alternate conceptions of introductory students are different from the alternate conceptions of upper-level students or if a peak is observed at one alternative conception for advanced students (as opposed to being spread across multiple alternate conceptions among introductory students). By "alternate conceptions" we mean physically incorrect ideas that sometimes arise out of common-sense notions. For the questions on which the upper-level EM students struggled, we identified potential reasons for why they were unable to answer those questions in those contexts after traditional instruction, both by asking upper-level undergraduate students to explain their answers to the CSEM questions in writing and by conducting individual interviews with graduate students in which they were asked to think aloud and reason why students would struggle with those questions if they did not know the correct answers.

**RESEARCH QUESTIONS**

To explore the progression in students' understanding and reasoning related to the CSEM, we investigated the following research questions while using the physics graduate student performance data on each CSEM question for benchmarking purposes:
- RQ1. On which CSEM questions do upper-level undergraduate students struggle after traditional instruction, where "struggle" means that less than two-thirds of the students answered the question correctly? Are there any patterns in student responses from introductory courses to upper-level courses for these questions?
- RQ2. What are some common reasons for why the upper-level students struggle on these questions in the post-test after instruction, i.e., what common alternate conceptions cause them to struggle on these questions?
- RQ3. On which CSEM questions do introductory students and upper-level students display different alternate conceptions in the post-test?
- RQ4. On which CSEM questions do the upper-level undergraduate students show moderate improvement or do very well in the post-test?
- RQ5. How does the performance of the first-year physics graduate students before graduate instruction in EM compare to the upper-level undergraduate students after instruction in post-test in different CSEM questions?

**METHODOLOGY**

All three levels of students involved in this study were from the same large public research university in the US. We used the CSEM, a validated survey, to investigate the progression in student understanding. We compare the EM conceptual knowledge of students in calculus-based introductory physics courses after traditional instruction with that of upper-level undergraduate EM course for physics majors enrolled in a mandatory course after traditional instruction. We benchmark the undergraduate student performance by comparing it with first-year physics graduate students enrolled in a first-semester teaching of physics course *before* they had taken their first-year physics graduate EM core course at the same university. In particular, we analyze the responses to the CSEM questions of first-year first-semester physics Ph.D. students who had not taken graduate level EM. These students had decided to pursue a Ph.D. in physics so they comprise a select group of students who had only taken an upper-level undergraduate EM course (i.e., had not taken any EM course beyond the upper-level undergraduate course) so a comparison of their responses with those of the upper-level undergraduates can provide useful benchmarks. Moreover, the majority of these graduate students were teaching introductory physics as teaching assistants (TAs) in addition to taking their own first-semester courses.

To gain an understanding regarding why undergraduate students may struggle with certain difficult questions on the CSEM, a separate set of 29 upper-level undergraduates in an EM course responded to the fourteen most challenging questions on the CSEM on which less than two thirds of the upper-level students in previous years provided the correct response and also provided their reasoning for why they thought that each answer they selected is correct. Additionally, we interviewed 16 physics Ph.D. students individually using a semi-structured think-aloud protocol to understand their views about the most common incorrect answers if students did not know the correct answers to those questions. In these interviews, we first asked students to think aloud and did not stop them when they explained their reasoning for what they thought were the correct answers or most common incorrect answers for each question. Only at the end we asked them for further clarifications of the points they had not made clear. Since graduate students are likely to have thought about these concepts both while learning these topics as undergraduates as well as while teaching introductory physics as TAs, they are likely to have good intuition about why certain CSEM questions are challenging for students. These Ph.D. students also served as TAs in a resource room where, among other duties, they helped introductory students with homework, often related to the content in introductory EM courses (in addition to mechanics and waves). In the first 11 interviews with the Ph.D. students, they were asked for the most common difficulties of students if they did not know the correct answers to the CSEM questions while in the last 5 interviews, they were also asked to specify if they thought there would be any difference between introductory and advanced students' most common difficulties. None of the graduate students thought that there would be a difference in the most common difficulties of the upper-level students and introductory students on each question.

With regard to the written administration of the survey, introductory students and upper-level undergraduate students took the CSEM before (pre) and after (post) instruction in relevant concepts, whereas data from the graduate students were collected once in their first-semester of first year before their graduate EM core course. The upper-level undergraduate EM course, which is required for physics majors, is usually taken by students in their second or third year. This course used Griffith's "Introduction to Electrodynamics" as the textbook and covered up to chapter 7 [45]. Data for this course were collected over three different years and averaged together since there were no significant differences across the years. These students took the pre-test within the first week of classes and the post-test in the last few weeks of classes. The introductory students in this study took a calculus-based physics course during their first year and were primarily engineering majors (~70%) with the rest being physics, chemistry, and mathematics majors. The graduate class in which the written CSEM survey was administered was a first-year (first semester) mandatory Teaching of Physics class which met for two hours each week for the entire semester and focused on preparing Ph.D. students for their job as teaching assistants (TAs) for introductory courses. Data were collected from four different years in this course for physics Ph.D. students and students took the CSEM prior to their graduate core EM course. All students used paper scantrons to bubble their responses to each question and had approximately 50 minutes, which was sufficient time for everyone to answer all the questions on CSEM.

In the introductory physics course, 456 students took the pre-test before instruction and 401 took the post-test after instruction. In the upper-level undergraduate course, 93 students took the pre-test and 85 students took the post-test. We present data from all introductory and upper-level students who took the pre- and post-tests because analysis using only matched data for pre- and post-tests yields very similar outcomes. There were 87 graduate students in the Teaching of Physics course who took the written CSEM spread over four consecutive years.

**RESULTS**

We find no practically significant difference between matched data (only students who took both the pre- and the post-test were included) and unmatched data (all students were included) in the introductory level as well as at the upper-level undergraduate courses. The overall average scores on the pre- and post-tests for matched and unmatched data for the introductory and upper-level students can be found in Table 1. Since the matched and unmatched scores are very similar, we report the detailed results for unmatched data below to make use of all data.

*RQ 1. On which CSEM questions do the upper-level undergraduate students struggle after traditional instruction, where "struggle" means that less than two-thirds of the students answered the question correctly? Are there any patterns in student responses from introductory courses to upper-level courses for these questions?*

To answer RQ1, we analyzed data to identify patterns in student responses across these questions with regard to the performance of both introductory and upper-level students. Fig. 1 shows the pre- and post-test performance of introductory and upper-level students for each question on the CSEM. There are 14 questions on which the performance of the upper-level students is below 2/3 (highlighted in yellow in Fig. 1: Q2, Q11, Q13, Q14, Q16, Q19, Q20, Q21, Q22, Q24, Q27, Q29, Q31, Q32). This is nearly half of the questions on the entire CSEM and indicates that the upper-level students are far from having developed a robust conceptual understanding of the basic EM concepts covered in the CSEM after traditional instruction in an upper-level EM course. Comparing the post-test performance of the upper-level students with the corresponding post-test performance of introductory students on these questions, we find that in nearly all of them (11 out of 14), introductory students' post-test performance after instruction is less than 50%. In fact, on **all** of the questions on which the performance of introductory students is below 50% on the post-test, the performance of the upper-level students after instruction is below two-thirds, indicating that the concepts covered in those questions are very challenging for even the upper-level students to grasp after traditional instruction, and instructors need to do more to help students develop a solid understanding of the underlying concepts using evidence-based approaches.

**Table 1.** Matched and unmatched data for the introductory and upper-level undergraduate courses.

|                    | Pre-test |           | Post-test |           |
|--------------------|----------|-----------|-----------|-----------|
|                    | Matched  | Unmatched | Matched   | Unmatched |
| Intro. Phys        | 38.1%    | 37.5%     | 53.9%     | 53.4%     |
|                    | N=353    | N=438     | N=353     | N=401     |
| Upper-level Phys.  | 63.9%    | 63.1%     | 75.3%     | 73.0%     |
|                    | N=69     | N=93      | N=69      | N=85      |

| CSEM Q# | Introductory students | | | Upper-level students | | |
|---|---|---|---|---|---|---|
| | Pre (N=438) | Post (N=401) | Effect size | Pre (N=93) | Post (N=85) | Effect size |
| 1 | 54% | 81% | 0.59 | 84% | 84% | -0.01 |
| 2 | 52% | 56% | 0.09 | 61% | 64% | 0.07 |
| 3 | 90% | 92% | 0.06 | 97% | 96% | -0.01 |
| 4 | 61% | 73% | 0.27 | 83% | 87% | 0.12 |
| 5 | 64% | 70% | 0.14 | 83% | 89% | 0.19 |
| 6 | 78% | 86% | 0.22 | 91% | 91% | -0.03 |
| 7 | 44% | 63% | 0.38 | 72% | 79% | 0.16 |
| 8 | 63% | 73% | 0.20 | 85% | 89% | 0.14 |
| 9 | 51% | 66% | 0.30 | 74% | 84% | 0.25 |
| 10 | 40% | 56% | 0.34 | 74% | 82% | 0.21 |
| 11 | 24% | 34% | 0.22 | 43% | 54% | 0.23 |
| 12 | 68% | 80% | 0.28 | 91% | 92% | 0.02 |
| 13 | 17% | 54% | 0.83 | 52% | 62% | 0.20 |
| 14 | 14% | 38% | 0.57 | 31% | 44% | 0.26 |
| 15 | 28% | 50% | 0.46 | 54% | 80% | 0.57 |
| 16 | 19% | 35% | 0.39 | 46% | 65% | 0.38 |
| 17 | 36% | 65% | 0.61 | 75% | 86% | 0.27 |
| 18 | 53% | 53% | -0.01 | 58% | 74% | 0.35 |
| 19 | 37% | 52% | 0.30 | 70% | 64% | -0.13 |
| 20 | 18% | 24% | 0.13 | 37% | 45% | 0.17 |
| 21 | 13% | 17% | 0.11 | 42% | 57% | 0.29 |
| 22 | 29% | 43% | 0.30 | 51% | 61% | 0.20 |
| 23 | 34% | 68% | 0.72 | 86% | 80% | -0.16 |
| 24 | 13% | 37% | 0.57 | 32% | 55% | 0.46 |
| 25 | 20% | 52% | 0.70 | 63% | 87% | 0.56 |
| 26 | 35% | 76% | 0.88 | 90% | 90% | 0.01 |
| 27 | 17% | 27% | 0.25 | 45% | 64% | 0.39 |
| 28 | 37% | 59% | 0.46 | 72% | 71% | -0.03 |
| 29 | 18% | 34% | 0.36 | 37% | 56% | 0.40 |
| 30 | 31% | 55% | 0.49 | 75% | 85% | 0.24 |
| 31 | 8% | 18% | 0.32 | 30% | 58% | 0.56 |
| 32 | 31% | 21% | -0.23 | 26% | 61% | 0.75 |

**Fig. 1.** Pre- and post-test performance of introductory and upper-level undergraduate students for each question on the CSEM along with effect sizes (Cohen's *d*) for the change from the pre-test to the post-test. Questions on which the post-test performance of upper-level students is lower than 2/3 are highlighted in yellow (14 questions) and questions on which the post-test performance is above 85% (10 questions) are highlighted in green.

*RQ2. What are some common reasons for why the upper-level students struggle on these questions in post-test after instruction, i.e., what common alternate conceptions cause them to struggle on these questions?*

We note that all the CSEM questions are provided in the appendix of Ref. [33]. In Table 2, we show the percentages of introductory students (post-test), upper-level students (post-test) and graduate students before instruction in graduate EM who selected each answer choice. In the appendix of this paper (Table 3), we show the same data as in Table 2 for the pre-test for introductory and upper-level students. We note that for the data reported in Fig. 1 and Tables 2 and 3, the standard error of proportion ranges from roughly 1.2% to 2.5% for introductory students and from roughly 1% to 5.4% for upper-level and graduate students (introductory: lowest 1.19%, highest 2.34%; upper-level: lowest 1.77%, highest: 5.41%; graduate students: lowest: 1.07%, highest: 5.31%).

**Table 2.** Percentages of introductory students and upper-level students who selected each answer choice on each question on the CSEM after instruction (post-test). The bold and underlined numbers represent the correct answer for each question number.

| Q# | Introductory students (post-test) | | | | | Upper-level students (post-test) | | | | | Graduate students | | | | |
|---|---|---|---|---|---|---|---|---|---|---|---|---|---|---|---|
|  | A | B | C | D | E | A | B | C | D | E | A | B | C | D | E |
| 1  | 3  | **_81_** | 14 | 2 | 1 | 2 | **_84_** | 11 | 4 | 0 | 2 | **_87_** | 9 | 1 | 0 |
| 2  | **_56_** | 15 | 13 | 10 | 5 | **_64_** | 6 | 6 | 17 | 7 | **_74_** | 2 | 5 | 14 | 5 |
| 3  | 5 | **_92_** | 3 | 0 | 0 | 0 | **_96_** | 2 | 1 | 0 | 0 | **_99_** | 1 | 0 | 0 |
| 4  | 2 | **_73_** | 17 | 8 | 0 | 0 | **_87_** | 12 | 1 | 0 | 1 | **_94_** | 5 | 0 | 0 |
| 5  | 14 | 4 | **_70_** | 8 | 3 | 9 | 0 | **_89_** | 1 | 0 | 5 | 0 | **_95_** | 0 | 0 |
| 6  | 5 | 6 | 3 | 1 | **_86_** | 2 | 4 | 1 | 2 | **_91_** | 3 | 1 | 2 | 0 | **_93_** |
| 7  | 5 | **_63_** | 28 | 4 | 1 | 1 | **_79_** | 19 | 1 | 0 | 1 | **_89_** | 9 | 0 | 1 |
| 8  | 1 | **_73_** | 8 | 16 | 3 | 0 | **_89_** | 0 | 6 | 5 | 1 | **_90_** | 1 | 5 | 3 |
| 9  | 6 | **_66_** | 10 | 17 | 2 | 5 | **_84_** | 1 | 11 | 0 | 1 | **_95_** | 1 | 2 | 0 |
| 10 | 8 | 24 | **_56_** | 5 | 6 | 2 | 12 | **_82_** | 2 | 1 | 2 | 5 | **_90_** | 1 | 2 |
| 11 | 37 | 3 | 12 | 13 | **_34_** | 34 | 0 | 6 | 6 | **_54_** | 18 | 1 | 5 | 1 | **_75_** |
| 12 | 6 | 8 | 5 | **_80_** | 2 | 5 | 1 | 2 | **_92_** | 0 | 2 | 0 | 1 | **_95_** | 1 |
| 13 | 26 | 17 | 2 | 2 | **_54_** | 19 | 17 | 0 | 2 | **_62_** | 10 | 13 | 0 | 0 | **_77_** |
| 14 | 34 | 14 | 5 | **_38_** | 9 | 29 | 16 | 1 | **_44_** | 9 | 35 | 10 | 0 | **_38_** | 16 |
| 15 | **_50_** | 12 | 32 | 4 | 2 | **_80_** | 7 | 12 | 0 | 1 | **_78_** | 9 | 10 | 0 | 2 |
| 16 | 24 | 14 | 16 | 11 | **_35_** | 17 | 4 | 10 | 5 | **_65_** | 5 | 6 | 1 | 2 | **_86_** |
| 17 | 2 | 8 | 18 | 6 | **_65_** | 0 | 6 | 6 | 2 | **_86_** | 0 | 2 | 5 | 1 | **_92_** |
| 18 | 0 | 6 | 11 | **_53_** | 31 | 0 | 0 | 1 | **_74_** | 25 | 0 | 2 | 3 | **_77_** | 17 |
| 19 | **_52_** | 23 | 14 | 6 | 7 | **_64_** | 22 | 4 | 2 | 8 | **_84_** | 12 | 0 | 0 | 5 |
| 20 | 17 | 25 | 30 | **_24_** | 5 | 7 | 20 | 21 | **_45_** | 7 | 1 | 10 | 17 | **_68_** | 3 |
| 21 | 14 | 30 | 27 | 13 | **_17_** | 4 | 13 | 16 | 11 | **_57_** | 2 | 11 | 15 | 2 | **_69_** |
| 22 | 15 | 9 | 31 | **_43_** | 2 | 2 | 2 | 34 | **_61_** | 0 | 2 | 6 | 21 | **_71_** | 0 |
| 23 | **_68_** | 10 | 6 | 11 | 4 | **_80_** | 2 | 6 | 11 | 1 | **_91_** | 1 | 1 | 6 | 1 |
| 24 | 3 | 23 | **_37_** | 34 | 3 | 0 | 13 | **_55_** | 32 | 0 | 0 | 5 | **_77_** | 17 | 1 |
| 25 | 11 | 12 | 17 | **_52_** | 9 | 6 | 0 | 5 | **_87_** | 2 | 3 | 0 | 2 | **_92_** | 2 |
| 26 | **_76_** | 7 | 7 | 7 | 3 | **_90_** | 1 | 4 | 1 | 4 | **_95_** | 1 | 1 | 0 | 2 |
| 27 | 21 | 21 | 11 | 20 | **_27_** | 7 | 11 | 7 | 12 | **_64_** | 7 | 2 | 3 | 2 | **_85_** |
| 28 | 7 | 5 | **_59_** | 3 | 26 | 1 | 2 | **_71_** | 0 | 26 | 2 | 0 | **_78_** | 1 | 18 |
| 29 | 21 | 25 | **_34_** | 19 | 2 | 11 | 21 | **_56_** | 12 | 0 | 3 | 25 | **_61_** | 6 | 5 |
| 30 | **_55_** | 11 | 15 | 17 | 3 | **_85_** | 2 | 7 | 4 | 2 | **_83_** | 2 | 1 | 8 | 6 |
| 31 | 11 | 25 | 33 | 13 | **_18_** | 12 | 8 | 5 | 17 | **_58_** | 14 | 1 | 2 | 14 | **_68_** |
| 32 | 39 | 21 | 10 | **_21_** | 10 | 26 | 4 | 4 | **_61_** | 6 | 24 | 1 | 12 | **_57_** | 6 |

On Q2, which asks what happens to a charge placed at a point P on the surface of an insulating sphere, on the post-test, 64% of upper-level students provided the correct response, and 17% incorrectly thought that after being placed there, most of the charge stays at point P, but some would spread over the sphere (answer choice D). Students who selected this incorrect answer choice and provided written explanations were not very clear, citing only that the sphere is insulating as a reason for their choice. However, in interviews, the physics graduate students who pondered over the most common incorrect answer were more articulate. For example, one of the graduate students stated that they expect that students would select this incorrect answer choice most commonly because students would know that an insulating sphere is different from a conducting sphere (Q1 was similar to question Q2, except the sphere was made of metal), but they don't fully understand the difference stating: "If they understand this is insulating material, they will choose D […] because they know something about insulating that it is not like the conducting, but they [may not know] that the charge will stay at the position [P]." In other words, students eliminate B and C because those answer choices mention that the charge "distributes evenly", which students are aware only happens if the object is conducting. Option choice E ("there is no excess charge left") is also not appealing (no upper-level students select it), and choice D is the only one that is clearly different from the conducting case (aside from the correct answer) that students could select.

On Q11, which asks what happens to the electric potential energy of a positive charge released from rest in a uniform electric field, the most common incorrect answer choice of upper-level students is that it stays the same. In interviews, graduate students often mentioned that students who did not know the correct answer may relate constant electric field with no change in electric potential (or electric potential energy). For example, one graduate student stated: "[Students would select A] because they will think that the field is uniform, so the potential is also the same, that is, constant everywhere." The written explanations of the upper-level students for this question also confirm that they used this type of reasoning. For example, two upper-level students provided the following reasoning for selecting choice A: "I believe EPE [electric potential energy] is proportional to E [electric field], so if one is constant, the other has to be." "Since the field is uniform, it will have the same electric potential throughout until there is a change in the magnitude or direction of the electric field."

Q13 and Q14 are the only two questions on the CSEM related to electrical shielding. On Q13 (see Fig. 2), the conducting sphere has been given a positive charge, and the question asks for the direction of the electric field at the center of the sphere. Here, 19% and 17% of upper-level students incorrectly noted that the field points to the left and right, respectively. When answering to the left, often, students who provided written or oral reasoning (during interviews) either motivated their answer choice by stating that the two objects repel each other because positive charges repel or appeared to ignore the conducting sphere altogether and state that electric field points away from positive charges. For example, one upper-level student stated: "The positive charge would cause the other positive charges to move away since they repel each other. This would cause the E field to point left" and another stated "Electric fields point away from positive charges" as the reason for selecting "to the left". When answering "to the right", students' explanation often noted that +Q outside will polarize the conducting sphere with the left of it positively charged and the right negatively charged (correct if the charge on the sphere is comparable to the +Q charge), and they concluded that the electric field at the center of the sphere would point to the right since the field points from positive to negative charges. In other words, they had some correct ideas, but carried them to an incorrect conclusion. For example, one student stated: "Negative charge will distribute itself around the surface by where +Q is, so the positive charge will distribute to the left. Field lines will point from positive to negative, so this would be from left to right." Interviewed graduate students who were asked about the most common incorrect answer if students did not know the correct answer often stated that answer choice A is the most common incorrect answer because students may not realize that the sphere has any effect. For example, one graduate student said: "Maybe someone would say leftward because they think of the positive being the source, so they think of it making a [electric field] line and the line is going out from the charge, and they think it's just going to go straight through the sphere." Another graduate student stated that he expects that some students may think that the charge spread over the sphere is uniform and because of this, it has no effect on the electric field in the middle of the sphere. Therefore, the electric field in the middle of the sphere is caused by the outside charge +Q and must point left.

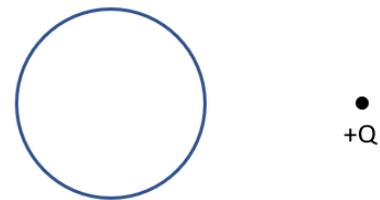

**Fig. 2.** Situation described on Q13.

Q14 is similar to Q13 except that the sphere is uncharged (but still conducting) and there is a charge +q at the middle of the spherical shell. The question asks about the forces experienced by the two charges. This is the question most resistant to change from the introductory level all the way to the graduate level so we provide the graduate student percentages as well: the percent of students who answered this question correctly was 38% (introductory), 44% (upper-level) and 38% (graduate), and the most common incorrect answer choice for all three groups of students was due to the application of Newton's 3$^{rd}$ law, namely that the two charges should feel equal and opposite forces (34% of introductory students, 29% of upper-level students, and 35% of graduate students). Answering that the forces are equal and opposite is consistent with the most common incorrect answer for Q13, namely not recognizing that the conducting sphere impacts the electric field inside it, making it zero. In written explanations for Q14, upper-level students sometimes explicitly mentioned the fact that the sphere is neutral for why the forces would be equal and opposite. For example, one upper-level student explained their reasoning: "The conducting metal sphere does not affect the force experienced by the charges, and [the forces] are always equal and opposite." There are two other alternate conceptions, although much less common: 1) that neither charge will feel a force; in other words, thinking that the conducting sphere shields both the inside and the outside (answer choice B, selected by 14% of introductory students, 16% of upper-level students, and 10% of graduate students) and 2) that both charges will feel a force, but those forces are different. Interviews with graduate students and written explanations from upper-level students suggest that some of them were still starting with Newton's 3rd law, but expecting that the sphere would reduce one of the forces, although not all the way to zero (answer choice E, selected by 9% of introductory students, 9% of upper-level students, and 16% of graduate students). For example, one student who selected E explained that "They both repel each other because they are both positive, but we don't know enough about them to determine the magnitude."

Q16 asks about the direction in which an electron would move if placed at a position on the $x$ axis where the electric potential is +10 V. Q16 was answered correctly (cannot predict the motion of the electron based on this information) by 65% of the upper-level students and the most common incorrect answer choice was that the electron would move to the left because it is negatively charged (17% of upper-level students). Written explanations of upper-level students and interviews with graduate students suggest that they often thought that students who provide this type of response might make other assumptions, e.g., that the electric field points to the right, or that the positive potential indicates that there is a flow of current towards the right (e.g., confusing electric potential with voltage or electric potential difference). For example, two upper-level students who selected choice A explained their responses as follows: "Unlike a positive test charge, an electron will move against the electric field." "The flow of a current from a potential of 10V will be right towards lower potential. Electrons move opposite the flow of the current."

Q19 provides students with vertical equipotential lines labeled as 10V, 20V, 30V, 40V, and 50V. A point A and a point B are on the 20V and 40V lines, respectively. Students are asked for the direction of the electric force that would act on a positive charge when placed at the two different points. The most common incorrect answer choice (22% of upper-level students) is to select the opposite direction to the correct answer. Interviews with graduate students and upper-level students' written reasoning suggests that those who selected this incorrect choice sometimes were thinking in terms of energy, electric potential and the motion of the charge. For example, some students stated that the charge will move in a direction of more positive electric potential because the charge is positive, one undergraduate said, "The object is positively charged so it will tend toward the higher potential difference."

Q20 provides the diagram shown in Fig. 3 and asks for the electric force acting on a proton placed at locations I and II. The answer choices are shown in Fig. 4. Roughly one fifth of upper-level students provided an incorrect answer for the magnitude (choice C), and a similar percentage got the direction backwards (choice B). Written explanations from upper-level students suggest that the difficulty with the magnitude is often because students didn't necessarily reason using the electric field, e.g., in the following statements: "I believe […] force is scaled with potential," or "with a higher potential there will be a higher force magnitude". With regard to the direction, the explanations suggest that students with this type of response often expected that the positive charges would move towards higher equipotential surfaces. Once again, this type of reasoning often did not include a discussion of the force explicitly, but the notion that positive charges move towards higher electric potential.

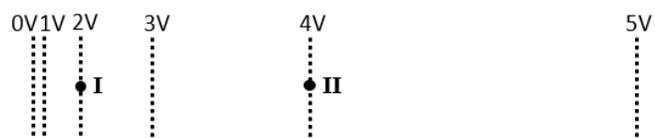

**Fig. 3.** Diagram provided for Q20.

**Fig. 4.** Answer choices for Q20.

Q21, Q22, Q27, and Q31 relate to the magnetic force on a charged particle in a magnetic field. Q21 asks what happens to a charge placed at rest in a uniform magnetic field, and the most common distractors for both groups of students (although for fewer upper-level students) are that the charge moves in a circle (options C and D, selected by 27% of upper-level students) and that it moves with constant acceleration (13% of upper-level students). Interviews with graduate students and written responses with explanations suggest that the upper-level students who answered that the charge moves in a circle (or accelerates in a circle) remembered that there was discussion of cyclical motion of charges in a magnetic field, but they did not realize that in order for this to happen, the velocity of the charged particle must be perpendicular to the magnetic field. For example, one student stated: "$F = q(v \times B)$ or something to that effect, which means that a charge placed in a magnetic field will experience a force perpendicular to the velocity," and another stated "I just remember this from [calculus-based physics 2] don't remember the explanation." It appears that students with this type of response had not developed a robust understanding of what causes the cyclical motion of a charged particle and only recalled that this was a common occurrence when discussing motion of charged particles in a magnetic field.

Q22 asks about the direction of a magnetic field that would cause a negatively charged particle to have a specific trajectory. The most common incorrect answer choice of both introductory and upper-level students was to choose the direction opposite to the correct direction with roughly comparable percentages (31% introductory, 34% upper-level students). Written explanations from upper-level students with this type of response suggests that sometimes the difficulty with this question is not due to being able to apply the right-hand rule correctly but rather due to not incorporating the sign of the charge. For example, one student explained: "Right hand rule when crossing v (cross) B, B must be into the page for the motion to be upwards."

Q27 provides the situation shown in Fig. 5 and states that the magnet on the left is three times as strong as the magnet on the right and that the positively charged particle is at rest. It asks for the magnetic force acting on the charged particle, making sure to emphasize the word "magnetic" by writing it in all caps and bold. Only 64% of the upper-level students answered this question correctly and those who answered it incorrectly were randomly spread among all of the incorrect answer choices (percentages range from 7% to 12%). For example, one upper-level student explained, "the left magnet would repel the charge and the right magnet would attract it." Written explanations suggest that upper-level students who selected any of the incorrect answer choices believed that the charge is either repelled or attracted to either magnet and depending on whether they thought that one force is larger than the other, they selected one of the incorrect answer choices. Thus, all these upper-level students expected that a charge at rest next to a magnet will feel a force.

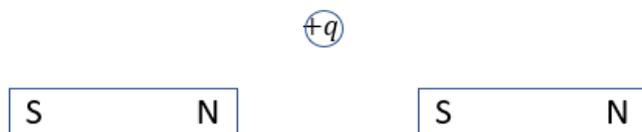

Fig. 5. Diagram provided for Q27.

Q31 is included in the CSEM in the group of questions relating to Faraday's law and Lenz's law. Even though this situation is often encountered when students learn about motional electromotive force (emf), the question only asks about the charge distribution on a neutral metal rod moving towards the right in an external magnetic field pointing out of the page. The correct answer can be obtained by using the right-hand rule: the electrons moving to the right in a magnetic field pointing out of the page would be deflected upwards; the top part of the metal bar will be negatively charged and the bottom would be positively charged (because the bar is neutral). On this question, upper-level students either selected the correct answer (58%) or used the right-hand rule incorrectly (17%). When providing reasoning, students who selected either of these answers reasoned using the right-hand rule. Based on these responses, it appears that these students generally knew the right-hand rule but were rusty using it. However, we point out that on this question, there are still one quarter of the upper-level students who did not realize that they could use the right-hand rule to answer it.

On two of the three questions focusing on Faraday's law, namely Q29 and Q32, the performance of the upper-level students was 56%, and 61%, respectively. Q29, see Fig. 6, provides four situations in which a bar magnet is next to a loop and asks students to identify all the situations in which there is an induced current in the loop. To answer the question correctly, students need to recognize that there is an induced emf both when there is relative motion between the loop and the magnet <u>and</u> when the area of the loop changes (such that the total magnetic flux passing through the loop changes). The most common incorrect answer choice for both introductory and upper-level students is due to the fact that students have difficulty recognizing that when the area of the loop changes, there must be an induced emf. The upper-level students who selected this answer and provided an explanation stated that in order for there to be an induced current, there must be motion between the loop and the magnet and/or the magnetic field must change. For example, one student stated: "[…] so long as the loop and magnet are moving perpendicular to each other relative to each other, they will generate a current due to the change in the magnetic field through the loop." In other words, students have difficulty recognizing that an emf is induced when the area of the loop changes, but can readily recognize that an emf is induced when there is relative motion between the magnet and the loop. This is confirmed when analyzing data for Q30, which is similar to Q29, except that it only included cases in which an induced emf is caused by relative motion between a source of a magnetic field (current-carrying wire) and a loop. This finding suggests that when assessing student understanding of Faraday's law, it is important to make use of both types of examples, and also include examples in which a rotation of a loop causes an induced emf.

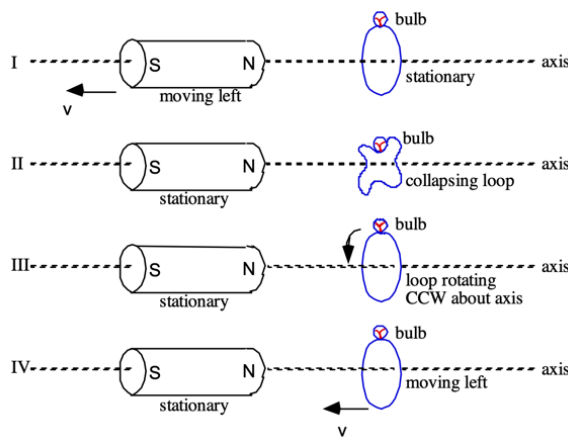

Fig. 6. Diagram provided for Q29.

Q32 (one of the most challenging questions on the CSEM) describes a situation in which two coils are close to one another. The first coil is connected to a power supply and an ammeter, and a graph of the relationship showing the current in the coil vs time is given (shown in Fig. 7). The second coil is connected to a voltmeter and students are asked for the time dependence of the voltmeter reading. A correct answer would recognize that the voltmeter reading is due to the induced emf in the second coil, which depends on the *rate of change* of magnetic flux through the coil, and thus would be constant and positive, then zero, then constant and negative, and then zero. In the most common incorrect answer choice selected by 26% of upper-level students, the induced voltage graph in the second coil looks the same as the current graph in the first coil. The upper-level students who provided written explanations and made this mistake often explicitly referred to $V = IR$ or mentioned that voltage is proportional to current in their reasoning. For example, one student stated: "V=IR, assuming the resistance is constant and even as the current in the second loop has to be induced, the voltage of the second loop will still be linearly related to the current and voltage of the first loop."

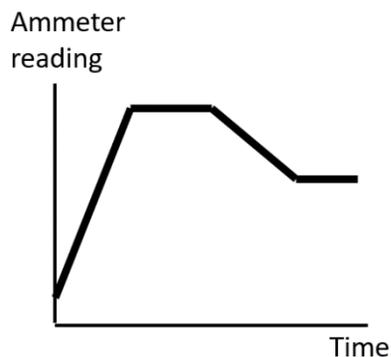

**Fig. 7.** Ammeter reading for coil connected to power supply on Q32

Q24 has two parallel wires carrying currents *I* and 3*I* in the same direction and the most common incorrect answer is that the current carrying wires feel repulsive forces (of the same magnitude) with 32% of upper-level students selecting this answer choice. Written explanations from students who selected this response often do not include a discussion of the right-hand rule, but rather an incorrectly memorized result that parallel wires carry current in the same direction repel. During the interviews, one graduate student who selected this as the most common incorrect answer stated that since two currents going into the paper would both be depicted by the same symbol (a circle with an x inside), this is somewhat similar to "like charges" so students may make a surface connection between the two situations and think that the current carrying wires with current in the same direction repel. Multiple upper-level students echoed this idea, one stating, "equal and opposite force still applies here and since the currents are in the same direction, they would act away from each other." This suggests that it is important for instructors to help students learn the difference between the cases involving current carrying wires and charges but also help students use two right hand rules to determine whether the wires repel or attract.

The other questions will be discussed in more detail when we answer RQ3, but the ones discussed above focus on the CSEM concepts that the upper-level students still struggled with after instruction in an upper-level EM course (less than two-thirds of the upper-level students provided the correct response).

If we group the questions on the CSEM based on the concepts they cover and try to identify the ones for which upper-level students generally have greater difficulty after traditional instruction, i.e., they have difficulty with most of the questions related to a particular concept, we find the most challenging concepts to be 1) Shielding (Q13 and Q14), and 2) Connection between electric field and electric potential (Q16, Q18, Q20), 3) Magnetic force on a charge in a magnetic field (Q21, Q22, Q27, Q31), and 4) Faraday's law/Lenz's law (Q29, Q30, and Q32).

With regard to shielding, the performance of upper-level students on both Q13 and Q14 is below two-thirds, and the performance on Q14 was not only the lowest on the entire survey, but also the singular question on which students at all levels (introductory, upper-level, graduate) showed comparable performance. As noted earlier, this is in large part due to students incorrectly using Newton's 3rd law in this situation when they contemplated the solution to this problem: it is very difficult to conceptualize that only one charge experiences a net force.

With regard to the connection between electric field and electric potential, most upper-level students appear to understand that the electric field is larger when the electric potential gradient changes faster (in space), but only when comparing electric fields at points in space where the electric potentials are the same (Q18, where post-test performance was 74%). If they are comparing electric fields at points where the electric potential is different (Q20), many upper-level students expected the electric field at the location with the larger electric potential to be larger even if the equipotential surfaces are farther apart at that location. On this question (Q20), upper-level students showed the second lowest performance on the survey. Similarly, if the only information available is the electric potential at a single point in space, many upper-level students believed that this is enough to determine the force on a charged particle placed at that location, often using reasoning focusing on the sign of the charge on the particle and the sign of the electric potential, e.g., thinking if the electric potential is positive at a certain location (for a 1D situation), a negative charge will move to the left.

With regard to the direction of the force on a charged particle in a magnetic field, students' performance on all four questions was below two-thirds. Students often thought that there must be a force on a charged particle even if the charge is stationary and seemed to sometimes use memorized results (e.g., a charge in a magnetic field moves in a circle) without recognizing the conditions of applicability (velocity must be perpendicular to the magnetic field). When a charge is close to a magnet (but stationary), many students expected that the charge is either attracted to or repelled by the magnet, often due to a confusion between electric and magnetic forces or thinking of north and south poles as charges. Lastly, when upper-level students needed to use the right-hand rule to determine the direction of the magnetic force, the most common incorrect answer was to select a direction opposite to the correct one. This suggests that instructors need to do more to help students recognize the difference between electric and magnetic forces and the necessity of motion of a charged particle in a magnetic field in order for there to be a magnetic force on it. In particular, with regard to these two concepts, upper-level student performance was not context specific as they performed poorly on all the questions related to them.

For the other two concepts (Faraday's law and Lenz's law), students generally struggled, but there were a few questions on which they performed well. For example, for Faraday's law, it appears that most upper-level students understand that when there is relative motion between a source of a magnetic field and a loop, an induced emf is created in the loop (Q30). However, many upper-level students had difficulty with the fact that an induced emf is created when the area of a loop changes (Q29), and also had difficulty conceptualizing a *changing* in a magnetic flux caused by a changing current (Q32). In the latter case, students are also distracted by Ohm's law due to the different quantities measured in the two loops (current in one, voltage in the other).

For the remaining concepts, students either showed context dependence (i.e., performing well in some contexts and poorly in others), or generally performed well in the majority of the contexts. One concept in which upper-level students performed well in all contexts was Coulomb's law (Q3, Q4, Q5, Q6, Q7, and Q8), where the post-test performance is greater than 85% in all questions except for one (Q7), on which the performance was 79%. Questions related to another concept on which upper-level students performed relatively well were the ones involving the connection between electric field and force (Q9, Q10, Q12, Q15, and Q19), on which their post-test performance was greater than 80% in all contexts with the exception of Q19, where the post-test performance was 64%.

### *RQ3. On which CSEM questions do introductory students and upper-level students display different alternate conceptions in the post-test?*

To answer RQ3, we identified the most common incorrect answer choices selected by both introductory and upper-level students for all questions and identified the questions in which the most common incorrect answer choices differed between introductory and upper-level student populations (also see Table 2). We find that in most cases, the most common incorrect answer choices of upper-level students were the same as for the introductory students, except the percentages were lower in the upper-level student group. In certain cases, the percentages of upper-level students were significantly lower than introductory students, to a point that the answer choice was selected by fewer than 10% or 5% of students. Thus, the alternate conceptions associated with those answer choices were not common among the upper-level students. In this section, we discuss primarily situations in which there is a difference between the alternate conceptions that are most common among one group compared to the other. When the alternate conceptions are the same as discussed in Maloney et al. [33] except less common or non-existent among the upper-level students, they are not discussed here. In other words, we focus on questions in which there are differences between upper-level and introductory students' alternate conceptions, but also on questions in which the percentages of introductory students and upper-level students with a specific alternate conception are similar, indicating a more challenging concept that instructors need to pay careful attention to with regard to helping students learn.

As shown in Table 2, on Q2, responses of the introductory students who answered the question incorrectly did not cluster around a particular incorrect answer choice, but among the upper-level students, 17% incorrectly expected that a short while later, most of the charge will still be at point P but some would spread over the sphere. Unlike the introductory students, this incorrect answer choice (D) was the most common among upper-level students with the other alternate choices at less than 10%.

On Q11, the most common alternate conception that the electric potential energy of a charge released in a uniform electric field is constant is roughly equally common among introductory (34%) and upper-level students (37%). Written explanations suggest that upper-level students who selected this answer choice expected that the field being uniform means that the electric potential energy does not change.

Q14, as mentioned earlier, is the most difficult question on the survey, and also the only question in which the percentages of both the correct and the most common incorrect answer are virtually indistinguishable among introductory, upper-level, and graduate students. It appears that Newton's 3$^{rd}$ law is very strong and students have a difficult time reconciling shielding with the notion that an action (force) must have an equal and opposite reaction force.

On Q18, the most common alternate conception of both introductory (31%) and upper-level students (25%) is that the electric field at a particular location only depends on the electric potential at that location (selecting that all fields have the same magnitude at point B in all three situations in Q18).

Using the right-hand rule incorrectly appears to still be common among upper-level students with roughly one third of students in both groups (31% of introductory and 34% of upper-level students) selecting the direction opposite to the correct direction in Q22. One student said, "You use left hand rule because it is a negative charge. When you line your pointer with the velocity and curl your fingers up with the direction it curves your thumb in pushed into the page. If you were supposed to use right hand rule then it would have been out of the page." Yet another student who had troubles using the right-hand rule correctly, even after stating that they knew that the charged particle was an electron, "The negative charge moves in the opposite direction a positive charge would. this can be found using the right-hand rule."

On Q24, in which two parallel wires carry currents $I$ and $3I$ in the same direction, students are asked for the magnitude and direction of forces on the wires. Roughly one third of both groups again selected the incorrect direction, and written explanations suggest that this mistake is often because the upper-level students used memorized knowledge to incorrectly infer that wires with currents flowing in the same direction repel. This difficulty is at least partly because students sometimes view wires with currents flowing in the same direction as analogous to like charges, which do repel.

On Q28, roughly one fourth of students (26% in both introductory and upper-level groups) incorrectly thought that the magnetic field from two loops will cancel halfway between the loops, both carrying current in the counter-clockwise direction. These students did not use the right-hand rule correctly in order to find the direction of the magnetic field produced by each loop, similar to other questions requiring use of the two right-hand rules.

In Q31, a neutral metal bar is moving to the right in a uniform magnetic field pointing out of the paper. The question asks about the resulting charge distribution on the metal bar. Compared to upper-level students, for introductory students the alternative conceptions related to charges separating in the same direction as the velocity (right or left) are more common (33% and 25% for introductory students, compared to 5% and 8% for upper-level students). However, for upper-level students, the alternate conceptions that charges do not separate (i.e., thinking that the metal bar remains neutral everywhere despite moving in the external magnetic field), or getting the direction wrong were more common (11% and 13% for introductory students, compared to 12% and 17% for the upper-level students).

On Q32 (setup discussed earlier), in both groups of students, the most common alternate conception is that the graph of the current as a function of time is the same as the graph of the voltage as a function of time for the other coil (as discussed earlier, upper-level students who provided reasoning for this type of response used in equation such as $V = IR$). However, this difficulty is less prevalent among upper-level students (26% for upper-level compared to 39% for introductory students). Furthermore, the alternate conception that the time dependence of the voltage as a function of time graph is the "opposite" or inverted compared to the current vs. time graph, i.e., when current increases, the voltage decreases is not prevalent among upper-level students (21% introductory, but only 4% upper-level). This suggests that, in general, upper-level students have learned not to misinterpret the meaning of "opposing" in Faraday's law, i.e., induced emf opposes the change in magnetic flux. However, many upper-level students did not recognize that Faraday's law should be used to answer Q32 and they used the much more familiar concept of Ohm's law.

### RQ4. On which CSEM questions do the upper-level undergraduate students show moderate improvement or do very well in the post-test?

To answer RQ4, we adopted a heuristic that 85% or more of the upper-level students answering a CSEM question correctly would correspond to good performance. This cutoff was chosen in part because for any multiple-choice test, it is possible for students to not read the information carefully or make mistakes due to simply being inattentive and not due to not understanding the concept. Moreover, there are some questions on the CSEM on which upper-level students performed very well at the end of their upper-level EM course, either because they have improved from the pre-test to post-test or because they already started with quite high performance on the pre-test. To quantify the improvement from the pre-test to post-test (see Fig. 1), we used effect size measured by Cohen's $d$. We focused on questions in which the effect size (measured by Cohen's $d$) for the improvement from the pre-test to the post-test was larger than 0.3. In particular, the general rule of thumb with regard to interpreting Cohen's $d$ is that a 'small' effect corresponds to an effect size of the order of 0.2, a 'medium' effect to 0.5 and a 'large' effect to 0.8 [46]. Of course, these effect sizes exist on a continuum and there are no clear demarcations as to where exactly a small effect size ends and medium effect size begins. Both these measures (85% score and effect size of 0.3 or higher) were used because if the pre-test performance of upper-level students is high on a question, there isn't much room for improvement, and it is unreasonable to expect large effect sizes. Also, analyzing the CSEM pre-test and post-test data for upper-level students, there appears to be a somewhat obvious break around this effect size of 0.3 where the next higher effect size for a CSEM question is much higher (e.g., there are four effect sizes between 0.25 and 0.3, but past 0.3, the next largest effect size is 0.35, then 0.38, 0.39, and 0.40 as shown in Fig. 1). The following questions fulfill these two criteria: Q15, Q16, Q18, Q24, Q25, Q27, Q29, Q31, Q32.

Q15 has the second highest improvement on the CSEM (from 54% to 80%) for upper-level students, which is primarily due to students recognizing that the direction of the electric force is opposite to the direction of the electric field for a negative charge. On Q16, while upper-level students still struggled at the end of the course (65% average), they showed marked improvement from the pre-test (effect size of 0.38). Student performance on Q16 suggests that it is very difficult even for many upper-level students to grasp that it is the *gradient* of the potential that is related to the electric field, and students struggled with the fact that the direction of the electric force on a charged particle cannot be found from information about the electric potential at just one point. Many students, even after traditional instruction in upper-level courses, thought that a single value of electric potential at a point can be used to determine what would happen to a charge placed at that location. On Q18, while students did show improvement (effect size 0.35), as mentioned earlier, a sizeable proportion of students thought that the electric field at a particular point depends on the electric potential at that point only and not on the change in electric potential. Q24 is very similar to Q18 in that students showed improvement, but as mentioned earlier, they retained a difficulty with similar percentages as introductory students. In the case of Q24, written responses from upper-level undergraduates suggest that many students did not explicitly use the two right-hand rules to determine that the parallel current carrying wires attract each other and incorrectly thought that the two wires that carry currents in the same direction would repel (sometimes based upon analogy with like charges which repel).

Q25 is the only question that both showed a marked improvement (effect size 0.56) and a good overall performance in the post-test (87%), showing that the upper-level students learned how the magnitude of the magnetic force depends on the angle between the velocity and the magnetic field. Q27 also showed reasonable improvement (effect size of 0.39), but the post-test performance was still less than two-thirds, and as discussed earlier, many upper-level students thought that a charge at rest next to two bar magnets would experience force due to the two magnets.

Q29, Q31 and Q32 are all from the group related to Faraday's law and Lenz's law and have been discussed earlier. The upper-level students showed some of the largest improvements from pre-test to post-test on these questions (the largest improvement on the entire CSEM was on Q32, effect size = 0.75) but they still struggled with these concepts at the end of the course, suggesting that the concept of induced emf opposing a *change* in *magnetic flux* through a coil via mutual induction is difficult to grasp even for upper-level students. On all these questions, students' pre-test performance was below 40%, showing that if the incoming knowledge of upper-level students related to these concepts is low, it is difficult for them to learn these concepts in an upper-level course taught traditionally.

With the exception of a single question, all the questions on which the upper-level students showed marked improvement were also questions on which their post-test performance was not higher than 85%. This shows how difficult these concepts are even for upper-level students to grasp. The questions on which the upper-level students performed well in the post-test (average greater than or equal to 85%) are Q3, Q4, Q5, Q6, Q8, Q12, Q17, Q25, Q26, Q30). Among these, with the exception of only one question (Q25, discussed earlier), the pre-test performance of upper-level students was already 75% or higher. Thus, it appears that overall, a traditionally taught upper-level EM course did not necessarily help students deepen their conceptual understanding of many of the basic concepts covered in CSEM.

**RQ5. How does the performance of the first-year physics graduate students before graduate instruction in EM compare to the upper-level undergraduate students after instruction in post-test at answering different CSEM questions?**

As shown in Table 2, the performance of first year Ph.D. students before their core graduate EM course was significantly better than the upper-level students on most CSEM questions. There were only three CSEM questions on which less than two thirds of the Ph.D. students provided the correct response. The most challenging question for the Ph.D. students is Q14 (related to electrical shielding), on which their average score was 38% (comparable to introductory and upper-level students on the post-test) even though on Q13, which was also related to electrical shielding, 77% of the Ph.D. students provided the correct response. The other two questions on which less than two-thirds of the Ph.D. students provided the correct response were Q29 (61% correct) and Q32 (57% correct) related to Faraday's and Lenz's laws. The significantly better performance of Ph.D. students compared to upper-level undergraduates before any instruction in EM at the graduate level may partly be due to the fact that this group consists of a select group of students and also because they were teaching assistants for introductory physics courses with some of them being TAs for introductory EM or they were helping introductory students with EM concepts in the physics resource room that all TAs at our institution do.

**SUMMARY AND CONCLUSION**

We find that many upper-level students struggled with introductory EM concepts on CSEM after traditional instruction in upper-level EM. The questions on the CSEM are conceptual in nature and upper-level students have difficulties with several conceptual aspects of EM covered in CSEM, even if they were able to carry out complicated calculations in their homework and exam problems similar to what was found in a study focusing on conceptual understanding of physics graduate students [47]. Additionally, we find that on only one question for which the upper-level students showed improvement, the Cohen's $d$ was larger than 0.3, and their post-test performance was over 85%. In other words, nearly all the questions on the post-test which showed a final performance of upper-level students larger than 85% started at 75% or higher at the beginning of the course in pre-test. The upper-level students often improved on questions on which they started very low in pre-test, but it was unlikely for them to improve beyond a certain level.

We find that on all the CSEM questions on which introductory students' performance was less than 50% in the post-test (after instruction), less than two-thirds of the upper-level students provided the correct response in the post-test. It appears that on questions in which their incoming knowledge is very low, it is very difficult to improve significantly after traditional instruction in an upper-level EM course on these concepts. We note that traditional upper-level EM teaching and assessment did not focus on the kinds of questions that are in the CSEM, although the course deals with the same concepts (chapters 1-7 of Griffith's textbook). Additionally, we find that the upper-level students often displayed the same types of difficulties on CSEM questions as introductory students, usually with lower percentages, but there wasn't a significant difference between the two groups on some of the questions. These findings suggest that in order to help students develop a robust knowledge structure, upper-level EM courses should use evidence-based approaches that integrate conceptual and quantitative aspects of EM. As Mazur noted [48, 49], students can become very adept at regurgitating solution patterns using memorized algorithms, but not be able to answer 'simpler' questions, e.g., comparing the brightness of different light bulbs in significantly simpler circuits. If instruction does not explicitly integrate both conceptual and quantitative problem solving in teaching and assessment, students can rely on algorithmic problem-solving approaches and get good grades despite lacking deep understanding of the underlying physics concepts.

In summary, these findings highlight that traditional upper-level EM courses are not effective in helping all students develop a solid grasp of concepts covered in CSEM. Instructors should not take for granted that students at the upper-level will make the conceptual and quantitative connections on their own if there is no intentional focus on this type of integration in their teaching and assessment [27]. They should focus on integrating both quantitative and conceptual aspects of EM in teaching and assessment using evidence-based approaches even in upper-level courses [50, 51]. We note that we previously conducted interviews with physics instructors who had taught traditional upper-level undergraduate and graduate core courses [27]. We found that some instructors incorrectly believe that learning physics concepts is easier for students than learning how to solve physics problems using "rigorous" mathematics. They believe that if non-science majors can learn physics concepts, science and engineering majors and particularly physics majors can learn physics concepts on their own even if there was no conceptual assessment in the course and there is not much use in wasting precious instructional time explicitly on concepts in advanced courses. Instructors with these types of beliefs often noted that they focus mainly on quantitative problem solving that will help students perform complex calculations. Moreover, interviews suggested that in upper-level or graduate courses, many instructors believe that students should have learned the concepts in the previous physics courses (e.g., in introductory courses for upper-level courses or in undergraduate courses for graduate-level courses) so their goal as instructors is mainly to focus on developing the "calculational" facility of students in the courses they are teaching instead of striving to appropriately integrate conceptual and quantitative understanding [27]. Some instructors also believe that students will focus on the physics concepts involved anyway in order to be able to do the calculations meaningfully so there is no need to reward them for conceptual understanding by asking them conceptual questions in assessment tasks. Other instructors claimed that they always mention important concepts involved before doing calculational problems or before doing complicated derivations in their classes. However, they only ask students to do calculations in assessment tasks that determine their course grade (these are often problems that many students learn to do without deep understanding of the basic underlying concepts). Our findings here in the context of CSEM suggest that many upper-level students did not develop a robust understanding of underlying concepts at the end of the course.

Also, even if students have had one round of exposure to concepts in previous courses, explicitly integrating conceptual and quantitative aspects of physics in instructional goals, instructional design and assessment of learning is critical for a majority of students to be motivated to focus on developing functional understanding. Students need to be supported to develop a robust knowledge structure of physics by solving a variety of integrated conceptual-quantitative problems that are appropriately scaffolded. Many researchers have pointed out that conceptual understanding and quantitative problem solving go hand in hand and that developing expertise in physics means becoming proficient in both. For example, Reif [52] discussed the importance of a good knowledge structure of physics and doing a conceptual analysis in the first step of problem solving. Similarly, the UMASS group [53] highlighted the role of conceptual understanding by focusing on qualitative problem-solving strategies. We note that some instructors in our earlier interviews [27] also claimed that they select calculational problems that have rich conceptual implications although they expect students to unpack those conceptual implications on their own when doing the calculation instead of explicitly integrating conceptual questions with those calculational problems in order to provide scaffolding support to make appropriate math-physics connections. Without such incentive and support, many students at all levels are not able to make such connections automatically and conceptual learning and development of a robust knowledge structure are compromised. Promisingly, instructors sometimes noted that they are happy to incorporate good conceptual questions or integrated conceptual-quantitative problems in their instruction, should such questions be available for them, but that they do not have the time to devote to creating these types of questions [28].

We hope that instructors of traditionally taught upper-level EM courses will use the findings of the research presented here as motivation to better integrate conceptual and quantitative aspects of EM in teaching and assessment. We also hope that physics education researchers will develop integrated conceptual-quantitative questions for upper-level EM courses that are cross-linked to commonly used textbooks to make it easy for the upper-level instructors to adapt such questions in their instructional design and assessment.

**ACKNOWLEDGMENTS**

This work was supported by Grant No. PHY-1806691 from the National Science Foundation. We would like to thank all students whose data were analyzed and Dr. Robert P. Devaty for his feedback on the manuscript.

# APPENDIX: PRE-TEST ANSWER CHOICES FOR INTRODUCTORY AND UPPER-LEVEL STUDENTS

For the interested reader, the frequencies of answer choices for each question on the CSEM for introductory and upper-level students in the pre-test are shown as percentages.

**Table 3.** Percentages of introductory students and upper-level students who selected each answer choice on each question on the CSEM before instruction (pre-test). The bold and underlined numbers represent the correct answer for each question number.

| CSEM Q# | Introductory Students (pre test) | | | | | Upper-Level Students (pre test) | | | | |
|---|---|---|---|---|---|---|---|---|---|---|
| | A | B | C | D | E | A | B | C | D | E |
| 1  | 4  | *54* | 29 | 11 | 2  | 2  | *85* | 9  | 3  | 1  |
| 2  | *52* | 12 | 8  | 17 | 11 | *61* | 12 | 1  | 15 | 11 |
| 3  | 2  | *90* | 3  | 4  | 0  | 2  | *97* | 1  | 0  | 0  |
| 4  | 2  | *61* | 27 | 9  | 1  | 1  | *83* | 14 | 2  | 0  |
| 5  | 19 | 7  | *64* | 6  | 3  | 14 | 1  | *83* | 2  | 0  |
| 6  | 6  | 10 | 5  | 1  | *78* | 4  | 2  | 2  | 0  | *91* |
| 7  | 7  | *44* | 43 | 5  | 1  | 3  | *72* | 22 | 2  | 1  |
| 8  | 2  | *63* | 9  | 24 | 3  | 1  | *85* | 0  | 12 | 2  |
| 9  | 8  | *51* | 15 | 22 | 4  | 3  | *74* | 6  | 16 | 1  |
| 10 | 11 | 20 | *40* | 10 | 20 | 0  | 18 | *74* | 3  | 6  |
| 11 | 40 | 12 | 12 | 12 | *24* | 35 | 1  | 9  | 12 | *43* |
| 12 | 16 | 7  | 7  | *68* | 2  | 0  | 1  | 7  | *91* | 0  |
| 13 | 57 | 19 | 5  | 2  | *17* | 33 | 15 | 0  | 0  | *52* |
| 14 | 57 | 9  | 3  | *14* | 16 | 46 | 9  | 3  | *31* | 12 |
| 15 | *28* | 17 | 46 | 6  | 3  | *54* | 16 | 24 | 3  | 3  |
| 16 | 17 | 20 | 22 | 22 | *19* | 16 | 22 | 8  | 9  | *46* |
| 17 | 4  | 19 | 32 | 9  | *36* | 1  | 8  | 12 | 4  | *75* |
| 18 | 4  | 6  | 14 | *53* | 23 | 0  | 2  | 4  | *58* | 36 |
| 19 | *37* | 27 | 18 | 11 | 7  | *70* | 19 | 5  | 3  | 2  |
| 20 | 22 | 28 | 28 | *18* | 3  | 17 | 13 | 32 | *37* | 1  |
| 21 | 20 | 33 | 22 | 12 | *13* | 7  | 22 | 16 | 13 | *42* |
| 22 | 26 | 14 | 20 | *29* | 10 | 7  | 11 | 32 | *51* | 0  |
| 23 | *34* | 24 | 22 | 11 | 8  | *86* | 3  | 3  | 5  | 2  |
| 24 | 3  | 38 | *13* | 38 | 8  | 2  | 26 | *32* | 37 | 3  |
| 25 | 17 | 18 | 37 | *20* | 7  | 18 | 6  | 8  | *63* | 4  |
| 26 | *35* | 28 | 9  | 23 | 4  | *90* | 3  | 2  | 3  | 1  |
| 27 | 13 | 37 | 12 | 21 | *17* | 11 | 22 | 8  | 15 | *45* |
| 28 | 9  | 16 | *37* | 7  | 31 | 2  | 2  | *72* | 1  | 23 |
| 29 | 31 | 24 | *18* | 21 | 6  | 16 | 28 | *37* | 18 | 0  |
| 30 | *31* | 17 | 31 | 15 | 7  | *76* | 7  | 5  | 7  | 5  |
| 31 | 7  | 27 | 41 | 17 | *8* | 18 | 18 | 13 | 21 | *31* |
| 32 | 18 | 21 | 16 | *31* | 14 | 39 | 9  | 13 | *27* | 11 |